\documentclass[12pt,a4paper]{article}
\usepackage{ulem}
\usepackage{graphicx,color}
\begin{document}

\begin{center}
\section*{Disk-Driven Rotating Bipolar Outflow in Orion Source~I}

Tomoya Hirota\footnotemark[1]$^{,}$\footnotemark[2]$^{,*}$, 
Masahiro N. Machida\footnotemark[3], 
Yuko Matsushita\footnotemark[3], 
Kazuhito Motogi\footnotemark[4]$^{,}$\footnotemark[5], 

Naoko Matsumoto\footnotemark[1]$^{,}$\footnotemark[6], 
Mi Kyoung Kim\footnotemark[7], 
Ross A. Burns\footnotemark[8], \& 
Mareki Honma\footnotemark[4]$^{,}$\footnotemark[9]

\end{center}

\footnotetext[1]{Mizusawa VLBI Observatory, National Astronomical Observatory of Japan, Osawa 2-21-1, Mitaka-shi, Tokyo 181-8588, Japan}
\footnotetext[2]{Department of Astronomical Sciences, SOKENDAI (The Graduate University for Advanced Studies), Osawa 2-21-1, Mitaka-shi, Tokyo 181-8588, Japan}
\footnotetext[3]{Department of Earth and Planetary Sciences, Faculty of Sciences, Kyushu University, Motooka 744, Nishi-ku, Fukuoka-shi, Fukuoka 819-0395, Japan}
\footnotetext[4]{Mizusawa VLBI Observatory, National Astronomical Observatory of Japan, Hoshigaoka2-12, Mizusawa-ku, Oshu-shi, Iwate 023-0861, Japan}
\footnotetext[5]{Graduate School of Sciences and Technology for Innovation, Yamaguchi University, Yoshida 1677-1, Yamaguchi-shi, Yamaguchi 753-8512, Japan}
\footnotetext[6]{The Research Institute for Time Studies, Yamaguchi University, Yoshida 1677-1, Yamaguchi-shi, Yamaguchi 753-8511, Japan}
\footnotetext[7]{Korea Astronomy and Space Science Institute, Hwaam-dong 61-1, Yuseong-gu, Daejeon, 305-348, Republic of Korea}
\footnotetext[8]{Joint Institute for VLBI in Europe, Postbus 2, 7990 AA, Dwingeloo, The Netherlands}
\footnotetext[9]{Department of Astronomical Sciences, SOKENDAI (The Graduate University for Advanced Studies), Hoshigaoka2-12, Mizusawa-ku, Oshu-shi, Iwate 023-0861, Japan}

{\bf{
One of the outstanding problems in star-formation theory concerns the transfer of angular momentum such that mass can accrete onto a newly born young stellar object (YSO). 
From a theoretical standpoint, outflows and jets are predicted to play an essential role in angular momentum transfer\cite{Blandford1982,Tomisaka2000,Pudritz2007,Machida2014} and their rotation motions have been reported for both low-\cite{Belloche2013} and high-mass\cite{Greenhill2013,Burns2015} YSOs. 
However, little quantitative discussion on outflow launching mechanisms have been presented for high-mass YSOs due to a lack of observational data. 
Here we present a clear signature of rotation in the bipolar outflow driven by Orion Source~I, a high-mass YSO candidate, using the Atacama Large Millimeter/Submillimeter Array (ALMA). 
A rotational transition of silicon monoxide (Si$^{18}$O) reveals a velocity gradient perpendicular to the outflow axis which is consistent with that of the circumstellar disk traced by a high-excitation water (H$_{2}$O) line. 
The launching radii and outward velocity of the outflow are estimated to be $>$10~au and 10~km~s$^{-1}$, respectively. 
These parameters rule out a possibility that the observed outflow is produced by entrainment of a high-velocity jet\cite{Arce2007}, and that contribution from stellar-wind\cite{Bouvier2014} or X-wind\cite{Shu1994} which have smaller launching radii are significant in the case of Source~I. 
Thus, present results provide a convincing evidence of a rotating outflow directly driven by the magneto-centrifugal disk wind launched by a high-mass YSO candidate\cite{Greenhill2013,Matthews2010}. 
}}

Orion Source~I is a radio emitting YSO candidate located in the Orion KL (Kleinmann-Low) region of the famous Orion nebula. 
As one of the nearest high-mass star-forming regions (418 pc from the Sun\cite{Kim2008}) it is an ideal target for high resolution observations. 
The $\sim$10$^{4}L_{\odot}$ central object\cite{Testi2010,Menten1995,Reid2007} drives a low-velocity ($\sim$18~km~s$^{-1}$) bipolar outflow along the northeast-southwest (NE-SW) direction at a 1000~au-scale, oriented at an almost edge-on view\cite{Greenhill2013,Plambeck2009,Zapata2012}. 
At the center of the outflow, vibrationally excited SiO masers trace a rotating outflow arising from the surface of a circumstellar disk with a radius of $\sim$50~au\cite{Matthews2010,Kim2008}, consistent with a magneto-centrifugal disk wind\cite{Greenhill2013,Matthews2010}. 
Recent high-angular resolution (100-200~au) ALMA observations of various molecular lines at high-excitation energies of 500--3500~K\cite{Hirota2012,Hirota2014,Hirota2016a,Plambeck2016} revealed a rotating hot molecular gas disk with an enclosed mass of (5$-$7)$M_{\odot}$.

\begin{figure}[th]
\begin{center}
\includegraphics[width=17cm]{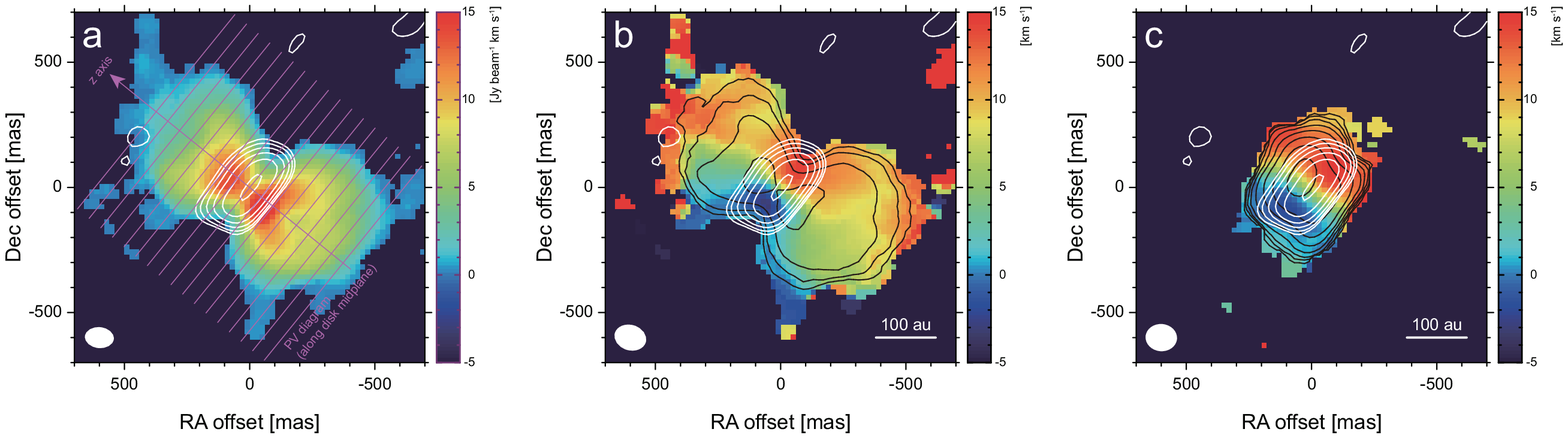}
\label{fig-moment}
\end{center}
{\bf{\noindent Figure 1:}} Moment maps of the observed lines and continuum emissions. 
(a) The 490~GHz continuum map (white contour) and moment 0 (integrated intensity; color) map of the 484~GHz Si$^{18}$O line. 
(b) The 490~GHz continuum map (white contour), moment 0 (black contour) and moment 1 (peak velocity; color) maps of the 484~GHz Si$^{18}$O line. 
(c) The 490~GHz continuum map (white contour), moment 0 (black contour) and moment 1 (color) maps of the 463~GHz H$_{2}$O line. 
Contour levels are 3,6,12,24, ...  times root-mean square (rms) noise levels, and the rms noise levels are 5~mJy~beam$^{-1}$, 481~mJy~beam$^{-1}$~km~s$^{-1}$, and 56~mJy~beam$^{-1}$~km~s$^{-1}$, respectively, for the continuum, moment 0 maps of Si$^{18}$O, and H$_{2}$O lines. 
Synthesized beam sizes are indicated at the bottom-left corner of each panel. 
In the panel (a), solid magenta lines indicate slices of the position-velocity (PV) diagram at 0, $\pm$60, $\pm$120, $\pm$180, $\pm$240, $\pm$300, $\pm$360, $\pm$420, and $\pm$480~mas from the disk midplane. 
The width of each slice is 60~mas. 
The interval of the slices parallel to the northwest-southeast direction is 60 mas corresponding to $\Delta z$=25~au at the distance of Orion~KL. 
The NE-SW outflow axis is defined as $z$ in Supplementary Figure \ref{fig-outflow}4. 
Position angle of the slice is determined from the Gaussian fitting of the continuum emission. 
\end{figure}

To reveal the interplay between disk accretion and outflow mechanisms in Source~I, we present observational results with ALMA at 110--120~mas, or 50~au resolution (Methods section). 
Figure \ref{fig-moment}1 shows integrated intensity and peak velocity maps of the 484.056~GHz Si$^{18}$O ($J$=12-11) and the 463.171~GHz H$_{2}$O ($\nu_{2}$=1, $J_{K_{a}, K_{c}}$=4$_{2,2}$-3$_{3,1}$) lines, along with the continuum emission at 490~GHz\cite{Hirota2016b}. 
The continuum emission has a deconvolved size of 90~au$\times$20~au extending perpendicular to the NE-SW bipolar outflow arising from the edge-on rotating disk\cite{Reid2007,Plambeck2016,Hirota2016b,Goddi2011}. 
The 484~GHz Si$^{18}$O ($J$=12-11) line clearly traces a NE-SW bipolar outflow, although the size of the outflow lobes, $\sim$200~au, are much smaller than that traced by the lower-energy transitions\cite{Plambeck2009,Zapata2012}. 
This is because the critical density required to excite the $J$=12-11 transition (6$\times$10$^{7}$~cm$^{-3}$) is higher than those of $J$=2-1 at 86~GHz (3$\times$10$^{5}$~cm$^{-3}$)\cite{Plambeck2009} and $J$=5-4 at 217~GHz (4$\times$10$^{6}$~cm$^{-3}$)\cite{Zapata2012} . 
It is also expected that the optically thin, less abundant Si$^{18}$O isotopologue ($^{16}$O/$^{18}$O$\sim$250)\cite{Tercero2011} selectively traces regions of higher density than the optically thick Si$^{16}$O lines. 
In contrast, the 463~GHz H$_{2}$O line shows a more compact structure than that of the Si$^{18}$O map. 
Because of its high excitation energy (2744~K), the 463~GHz H$_{2}$O line would be emitted from the hot molecular gas disk and base of the outflow closer to the central YSO\cite{Hirota2014,Plambeck2016} than the 484~GHz Si$^{18}$O line. 

\begin{figure}[th]
\begin{center}
\includegraphics[width=17cm]{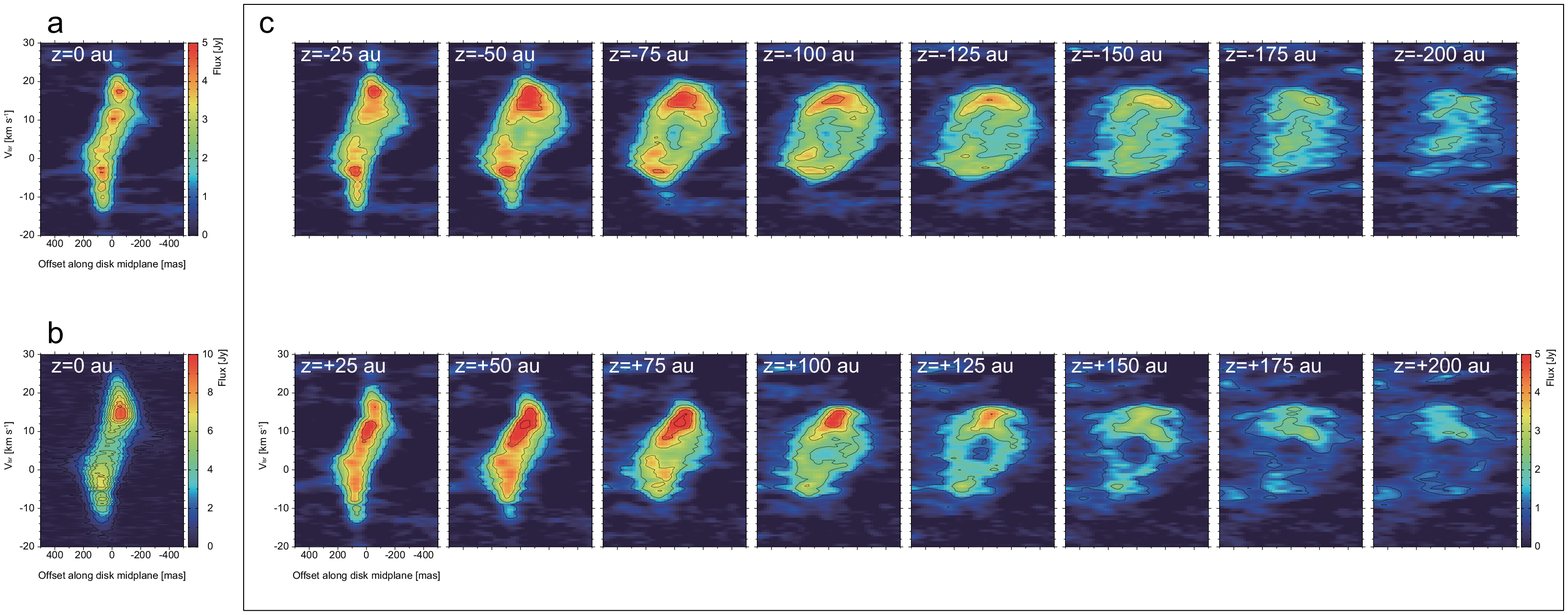}
\label{fig-pvmap}
\end{center}
{\bf{\noindent Figure 2:}} Position-velocity (PV) diagrams parallel to the disk midplane. 
(a) The 484~GHz Si$^{18}$O line at the offset $z$=0~au (midplane). 
Contour levels are 3,6,12,24, ...  times rms noise level of 358~mJy. 
(b) The 463~GHz H$_{2}$O line at the offset $z$=0~au (midplane). 
Contour levels are 3,6,12,24, ...  times rms noise level of 166~mJy. 
(c) The same as (a) but at different offset positions from $z$=$\pm$25~au to $\pm$200~au with an interval of 25~au as shown in Figure \ref{fig-moment}1(a). 
Contour levels are the same as panel (a) but with the different rms noise levels of 310--390~mJy for each position offset. 
\end{figure}

Both the 484~GHz Si$^{18}$O and 463~GHz H$_{2}$O maps clearly show velocity gradients along the disk plane (Figures \ref{fig-moment}1(b)(c)). 
Velocity gradients across the outflow or jet axis of other sources have been detected and have been attributed to rotation motions\cite{Belloche2013}. 
Alternative scenarios have also been suggested to explain the observed gradients, such as precessing motion, interaction with a warped disk, or two unresolved jets. 
However, we could not see any hints of wiggled or S-shaped point symmetric structure, disagreement of velocity structure in opposing outflow lobes or between disk and outflow, unlike previous studies. 
In contrast, we spatially resolve the velocity gradients in both lobes of the molecular outflow traced by the Si$^{18}$O emission, which are similar in the observed H$_{2}$O emission (which traces the disk) (Figure \ref{fig-pvmap}2 and Supplementary Figure \ref{fig-chmap}5). 
Furthermore, all the observed parameters/properties can be explained consistently via the rotation motions of the disk and outflow as discussed below. 
Thus, we conclude that outflow rotation is the best explanation for the velocity gradients in Source~I. 

The signature of rotation is clearly demonstrated in the position-velocity (PV) diagrams of the 484~GHz Si$^{18}$O and 463~GHz H$_{2}$O emission in the disk midplane (Figures \ref{fig-pvmap}2(a)(b)). 
These are analogous to those of the high excitation lines detected by the recent ALMA observations\cite{Hirota2014,Plambeck2016}. 
The PV maps exhibit a signature of a central hole (Supplementary Figure \ref{fig-model}1(b)) suggesting that emission from the outer parts of the outflow is dominant, and/or that high opacity continuum emission could obscure the innermost region in the midplane\cite{Hirota2014,Plambeck2016,Hirota2016b}. 
The inner and outer radii of the disk, and rotation velocities at these positions are derived from the Si$^{18}$O map, $R_{in}$($z$=0)=24$\pm$8~au, $R_{out}$($z$=0)=76$\pm$4~au, $v_{\phi}$($z$=0,$r$=$R_{in}$)=17.9$\pm$0.6~km~s$^{-1}$, and $v_{\phi}$($z$=0,$r$=$R_{out}$)=7.0$\pm$0.4~km~s$^{-1}$ with 1$\sigma$ errors (Methods section). 
While the above values can be explained by a rotation curve conserving angular momentum, $v_{\phi} \propto r^{-1}$, the PV diagrams could not distinguish Keplerian rotation, $v_{\phi} \propto r^{-0.5}$ due to insufficient resolution. 
Assuming Keplerian rotation, the enclosed mass is estimated to be 8.7$\pm$0.6$M_{\odot}$ (1$\sigma$) from the maximum rotation velocity at the inner radius. 
The result is consistent with the mass estimated from SiO maser observations\cite{Matthews2010}. 
However, the shallower velocity gradient around the systemic velocity (5~km~s$^{-1}$) agrees with a smaller enclosed mass\cite{Hirota2014,Plambeck2016}. 
It is also likely that the inner radii and resultant enclosed mass could be overestimated due to insufficient angular resolution. 

We made slices for PV diagrams of the 484~GHz Si$^{18}$O line at different offset positions from the disk midplane, $z$ (Figure \ref{fig-moment}1(a)). 
The PV diagrams at $z\sim\pm$25~au show velocity gradients similar to that along the midplane (Figures \ref{fig-pvmap}2(a)(c)). 
Because of the beam size of 50~au which is larger than the 25~au width and interval of the PV diagram slices, the data could be affected by contamination from the disk midplane. 
However, elliptical structures in the PV diagram with slight inclination become more clear at $z=\pm$75~au or larger distances from the midplane. 
These can be interpreted as rotation with radial expansion perpendicular to the outflow axis (Supplementary Figure \ref{fig-model}1(d)). 
Such a clear and continuous structural change in the PV diagrams of the outflow directly connecting from the disk with rotation and radial expansion is revealed for the first time in the present study. 
It is a major difference from previous maser observations which could measure proper motions at higher angular resolution but with discrete sampling of spatial structures\cite{Greenhill2013,Burns2015}. 
These findings are facilitated by utilizing high excitation thermal emission from the H$_{2}$O and Si$^{18}$O lines.

\begin{figure}[th]
\begin{center}
\includegraphics[width=17cm]{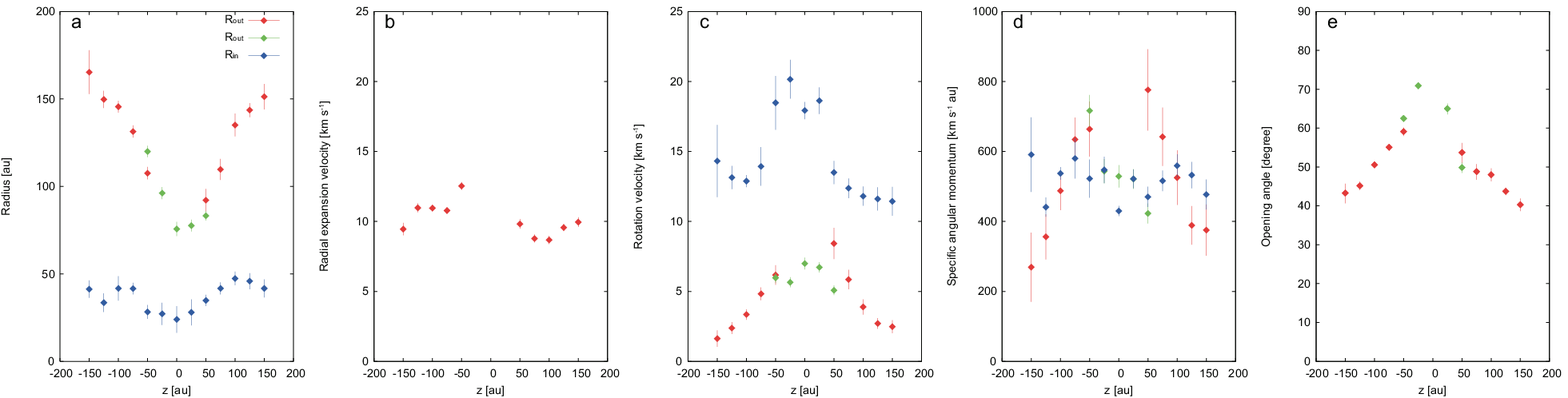}
\label{fig-param}
\end{center}
{\bf{\noindent Figure 3:}} Derived outflow parameters from PV diagrams of the 484~GHz Si$^{18}$O line. 
All the error bars (1$\sigma$) are estimated from error propagation from the fitting on PV diagrams. 
(a) Outer and inner radii of the outflow, $R_{out}$ and $R_{in}$,  
(b) radial expansion velocity perpendicular to the outflow axis at the outer radius, $v_{r}(R_{out})$, 
(c) rotation velocity at the outer and inner radii, $v_{\phi}(R_{out})$ and $v_{\phi}(R_{in})$, 
(d) specific angular momentum at the outer and inner radii, $J(R_{out})$ and $J(R_{in})$, and 
(e) opening angle of the outflow $\theta$ (Supplementary Figure \ref{fig-outflow}4). 
The fitting results to the linear velocity gradients across the outer radius (green) and  inner radius (blue) are plotted for the offset within $|z|\le$50~au from the midplane. 
The fitting results to the elliptical function (red) are plotted for the offset of $|z|\ge$50~au or larger from the midplane. 
We note that both linear and elliptical function fitting are done at $z$=$\pm$50~au for comparison (see Methods section). 
\end{figure}

PV diagrams along the outflow axis reveal the detailed structure of the outflow as a function of the distance from the disk midplane (Methods section and Figure \ref{fig-param}3). 
The inner and outer radii of the outflow, $R_{in}$ and $R_{out}$, increase at larger offset from the midplane (Figure \ref{fig-param}3(a)). 
The radial expansion velocity perpendicular to the outflow axis, $v_{r}$($r$=$R_{out}$)$\sim$10~km~s$^{-1}$ shows no significant trend as a function of the distance from the disk midplane (Figure \ref{fig-param}3(b)). 
In contrast, rotation velocities at both of the inner $v_{\phi}$($r$=$R_{in}$) and outer $v_{\phi}$($r$=$R_{out}$) radii are almost constant close to the midplane within $|z|\le$50~au, while those at the outer radius slow down significantly at larger offsets from the midplane (Figure \ref{fig-param}3(c)). 
The specific angular momentum, $J$=$v_{\phi} r$, of the outflow is almost constant within $|z|\le 50$~au, 
particularly for the inner radius $R_{in}$, while there is still a significant amount of specific angular momentum at the larger distance from the disk midplane (Figure \ref{fig-param}3(d)). 
Given the enclosed mass of 8.7$M_{\odot}$ and the conservation of angular momentum, the specific angular momentum of 400-600~km~s$^{-1}$~au corresponds to centrifugal radii of 21-47~au, where the gravitational and centrifugal forces balance. 
The derived centrifugal radii agree with that of the observed disk size (Figure \ref{fig-param}3(a)), implying that the outflow mainly originates in a region close to the derived centrifugal radii.
Such a rotating outflow launched in the disk outer radii is most likely explained by the magneto-centrifugal disk wind mechanism, adding strong support to previous claims\cite{Greenhill2013,Matthews2010}. 
Similarly, the launch radii of magneto-centrifugal disk winds in low-mass YSOs are also comparable to their disk radii, as seen in recent ALMA observations of TMC-1A\cite{Bjerkeli2016}.
The centrifugal radii and specific angular momentum in Source~I are larger than those of TMC-1A due to its larger stellar mass. 
We note that the specific angular momentum at the outer radius $R_{out}$ shows a slightly decreasing trend as the outflow moves away from the disk midplane (Figure \ref{fig-param}3(d)), which contrasts the trend seen in TMC 1A\cite{Bjerkeli2016}. 
This is probably due to the dissipation of angular momentum from the outflow to the ambient gas or the contribution from the swept-up ambient gas with smaller angular momentum by the outflow in Source~I. 
The opening angle of the outflow (Supplementary Figure \ref{fig-outflow}4) has a maximum value of 70~degrees close to the midplane $z$=$\pm$25~au (Figure \ref{fig-param}3(e)). 
Such a wide opening angle close to the launching point and collimation at a larger distance, which could be achieved by hoop stress\cite{Pudritz2007}, are also indicative of a magneto-centrifugal disk wind. 

The dynamical timescale of the outflow traced by the Si$^{18}$O line is estimated from the outer radius and radial expansion velocity, $t_{dyn}$=$R_{out}/v_{r}$=70$\pm$4~yr (1$\sigma$) at $z$=$\pm$150~au. 
It should be noted that the Si$^{18}$O line traces only the central part of the larger NE-SW bipolar outflow with a much longer dynamical timescale\cite{Plambeck2009}. 
Using this dynamical time, the outward velocity along the outflow axis, $v_{z}$, is estimated to be 10~km~s$^{-1}$. 
This is close to the three-dimensional velocity of the low-velocity "18~km~s$^{-1}$" outflow which is measured by proper motions of H$_{2}$O masers\cite{Greenhill2013}, but much lower than those observed in optical jets driven by T-Tauri stars\cite{Belloche2013}. 
The three-dimensional velocity of 18~km~s$^{-1}$ is consistent with the maximum rotation velocity in the disk midplane, 17.9~km~s$^{-1}$ at $R_{in}$($z$=0)=24~au. 
The similarity lends further support to our inferred outflow launching radius. 
Using the derived enclosed mass (8.7$M_{\odot}$) and outward velocity along the outflow (10~km~s$^{-1}$), we can also derive the launching radii of the outflow by using the magneto-centrifugal disk wind model\cite{Anderson2003} (see their equation (4)). 
The calculated values, 5-25~au, are comparable with or slightly smaller than the centrifugal radii as estimated above (21-47~au). 
If we instead adopt an outward velocity of 18~km~s$^{-1}$ from maser kinematics\cite{Greenhill2013}, the estimate launching radii become smaller by 10-25\%. 
Alternative outflow mechanisms include entrainment of ambient gas by a high-velocity ($>$100~km~s$^{-1}$) primary jet\cite{Arce2007}. 
However, the kinematics discussed above are better explained by an outflow directly driven by a disk wind, rather than entrainment by a narrow high-velocity jet which has never been identified toward Source~I. 
The above estimated launching radius suggests that contribution from other mechanisms (stellar-wind\cite{Bouvier2014} or X-wind\cite{Shu1994} which are launched from a close vicinity to the central YSO) would be small in the case of Source~I. 

In Orion~KL, high velocity ($\sim$100~km~s$^{-1}$) explosive outflows have been identified in infrared and radio observations which are thought to have been triggered by a dynamical encounter event 500~years ago\cite{Bally2011,Zapata2009}. 
It is predicted that Source~I is a binary system newly formed during the encounter event with a total mass and binary separation of 20~M$_{\odot}$ and $<$10~au, respectively\cite{Goddi2011,Bally2011} (although the mass, luminosity, and temperature of the central star or binary system is still controversial\cite{Testi2010,Reid2007,Hirota2014,Plambeck2016,Goddi2011,Chatterjee2012}). 
Even if we assume a higher mass (20$M_{\odot}$) and a binary system, the corresponding centrifugal radii, 9-20~au, would not conflict with the magneto-centrifugal disk wind driven by a circumbinary disk\cite{Machida2009}. 
Thus, our results provide a compelling observational signature of a rotating outflow driven by the magneto-centrifugal disk wind launched by a high-mass YSO candidate, Source~I. 
This picture is analogous to those of low-mass YSO cases\cite{Bjerkeli2016}, in which mass accretion occurs through disks by carrying away significant amount of angular momentum by the magneto-centrifugal disk wind, as theoretical models predict\cite{Blandford1982,Tomisaka2000,Pudritz2007,Machida2014}. 
The present study reveals not only the outflow driving mechanism but also proves a possible scenario of mass accretion and angular momentum transfer processes in high-mass star-formation via the disk accretion.


\clearpage

\clearpage

\section*{Methods}

\subsection*{ALMA observations and data analysis}

Observations of the submillimeter Si$^{18}$O and H$_{2}$O lines were carried out on July 27 (490~GHz) and August 27 (460~GHz), 2015 with the Atacama Large Millimeter/Submillimeter Array (ALMA) as one of the science projects in cycle~2 (2013.1.00048). 
The target source was the radio Source~I in Orion~KL (Orion Source~I) and the tracking center position was RA(J2000)=05h35m14.512s, Decl(J2000)=$-$05d22'30.57". 
The total on-source integration time of the target source was 410~seconds and 315~seconds for 463~GHz and 484~GHz, respectively. 
The vibrationally ground Si$^{18}$O line at 484.056~GHz ($J$=12-11, $E_{l}$=128~K\cite{Muller2005}) and the vibrationally excited H$_{2}$O line at 463.171~GHz ($\nu_{2}$=1, $J_{K_{a}, K_{c}}$=4$_{2,2}$-3$_{3,1}$, $E_{l}$=2744~K\cite{Pickett1998}) in the ALMA band~8 were selected for the present paper. 
A primary flux calibrator, band-pass calibrator, and secondary gain calibrator were J0423$-$013, J0522$-$3627, and J0607$-$0834, respectively. 
The array consisted of 41 and 40 antennas with a diameter of 12~m each in the configuration with the maximum baseline length of 1466.2~m and 1574.4~m for the 484~GHz and 463~GHz, respectively. 
The primary beam size was 12.05 arcseconds. 
The median system noise temperature was around 400~K at both frequency bands. 
The ALMA correlator provided the four spectral windows with the total bandwidth of 
1875~MHz and 937.5~MHz for the 484~GHz Si$^{18}$O and 463~GHz H$_{2}$O lines, respectively. 
The channel spacing of spectrometers were 976.562~kHz and 488.281~kHz for the 484~GHz Si$^{18}$O and 463~GHz H$_{2}$O lines, respectively, which corresponds to the velocity spacing of 0.6~km~s$^{-1}$ and 0.3~km~s$^{-1}$, respectively . 

\begin{figure}[th]
\begin{center}
\includegraphics[width=17cm]{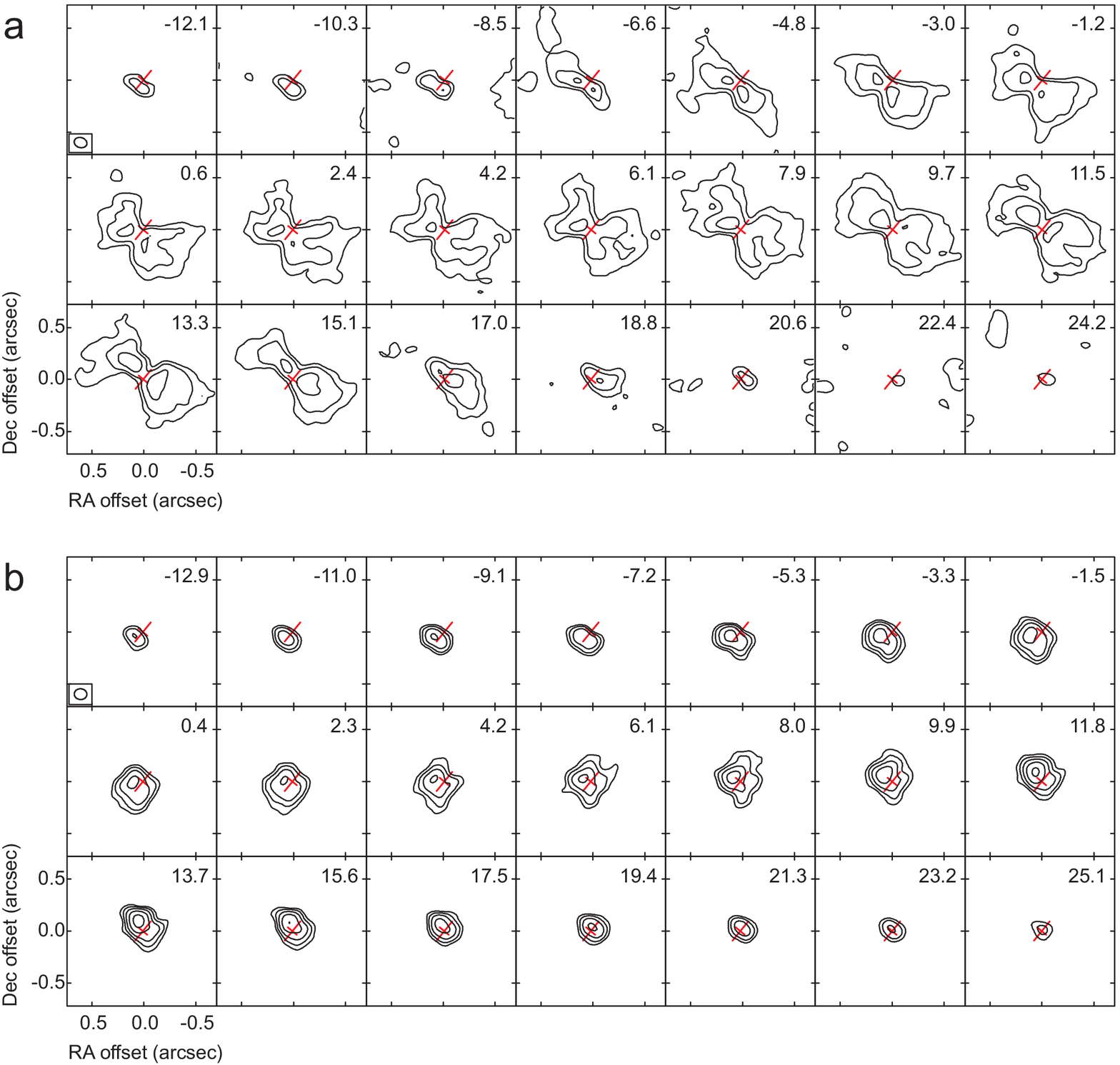}
\label{fig-chmap}
\end{center}
{\bf{Supplementary Figure 5:}} Channel maps of (a) the 484~GHz Si$^{18}$O and (b) 463~GHz H$_{2}$O lines. 
Contour levels are 10\%, 30\%, 50\%, 70\%, and 90\% of the peak intensity of 0.92~Jy~beam$^{-1}$ and 2.2~Jy~beam$^{-1}$ for the 484~GHz Si$^{18}$O and 463~GHz H$_{2}$O lines, respectively. 
The rms noise levels are 0.048~Jy~beam$^{-1}$ and 0.033~Jy~beam$^{-1}$ for the 484~GHz Si$^{18}$O and 463~GHz H$_{2}$O lines, respectively. 
A red cross in each panel indicates the position and size (major and minor axes) of the continuum emission. 
\end{figure}

The synthesis imaging and self-calibration were done with the CASA software package. 
We employed the calibrated data delivered by the East-Asia ALMA Regional Center (EA-ARC). 
First, visibility data were separated into spectral lines and continuum emissions by setting the line free channels with the CASA task {\tt{uvcontsub}}. 
Next, both phase and amplitude self-calibration was done with the continuum emission of Source~I by integrating over all the channels using CASA tasks {\tt{clean}} and {\tt{gaincal}}. 
Because of calibration problems in the delivered data due to the large atmospheric opacity around 487~GHz, we used only three spectral windows among four ALMA basebands to make a continuum image at 490~GHz\cite{Hirota2016b}. 
The phase and amplitude solutions of self-calibration were applied to all the spectral channels including the target lines by the CASA task {\tt{applycal}}. 
The uniform-weighted synthesized images were made for both 484~GHz Si$^{18}$O and 463~GHz H$_{2}$O lines by using the CASA task {\tt{clean}} providing velocity channel maps (image cubes; Supplementary Figure \ref{fig-chmap}5). 
The resultant synthesized beam size were 124~mas$\times$97~mas with the position angle of 73.3~degrees at 484~GHz and 121~mas$\times$104~mas with the position angle of 85.2~degrees at 463~GHz. 
The maps of the 484~GHz Si$^{18}$O and 463~GHz H$_{2}$O lines were registered with that of the 460~GHz and 490~GHz continuum maps, respectively\cite{Hirota2016b}. 
The moment maps were created by the CASA task {\tt{immoments}}. 
Some of the analysis and plotting were done with the AIPS software package. 
We made position-velocity (PV) diagrams by using AIPS in a standard manner. 

The produced PV diagrams were analyzed to derive the physical parameters by fitting the images. 
We found that the observed PV diagrams show different velocity structures as seen in Figure \ref{fig-pvmap}2(c). 
These PV diagrams can be interpreted as a combination of rotation and radial expansion motions perpendicular to the outflow axis. 
Schematic model PV diagrams are shown in Supplementary Figure \ref{fig-model}1. 
If the slice of the rotating outflow or disk has a ring-like structure with a central hole, one can see a linear velocity gradient without high-velocity components associated with the inner part. 
If the gas radially expands outward with significant rotation, the PV diagram becomes elliptical morphology with velocity gradients. 

To derive the radius of the outflow, rotation velocity, and radial expansion velocity, we fitted the linear function or elliptical function to the PV diagrams (Supplementary Figure \ref{fig-model}1). 
The examples are shown in Supplementary Figures \ref{fig-sample}2 and \ref{fig-sample2}3. 
We determined the peak position and those of the full-width at half maximum (FWHM) for each velocity channels by Gaussian fittings to the intensity profile (Supplementary Figures \ref{fig-sample}2(c) and \ref{fig-sample2}3(c)). 
The linear velocity gradient is applied to the PV diagrams within $z$=$\pm$50~au or smaller offset from the disk midplane. 
For PV diagrams with the elliptical shape, we employed two Gaussians to determine the double-peaked intensity profiles if there are two peak positions at the same velocity channel (Supplementary Figure \ref{fig-sample2}3(c)). 
The elliptical function fitting is conducted for the PV diagrams at the offset of $z$=$\pm$50~au or larger distance from the disk midplane. 
For comparison, we applied both fitting procedure at $z$=$\pm$50~au. 
To derive linear velocity gradients across the disk inner radius, we carried out fitting to use the highest velocity components with the width of 6~channels (3~km~s$^{-1}$) as shown in Supplementary Figures \ref{fig-sample}2(a) and \ref{fig-sample2}3(a). 
On the other hand, linear velocity gradients across the disk outer radius are obtained by fitting the linear function using the outermost part corresponding to the FWHM positions with the velocity width of 3~km~s$^{-1}$ (or 6~channels), as shown in Supplementary Figure \ref{fig-sample}2(b). 
To fit the elliptical-shaped PV diagrams, we only used the outer boundary corresponding to the FWHM positions in the fitting (Supplementary Figure \ref{fig-sample2}3(b)). 
The quoted uncertainties for the peak positions and half maximum positions are estimated from the fitting errors to the PV diagrams with 1$\sigma$ level. 

Using the above fitting results, we estimate the outflow parameters such as the inner and outer radii, $R_{in}$ and $R_{out}$, the radial expansion velocity perpendicular to the outflow axis, $v_{r}$, and the rotation velocity, $v_{\phi}$, as a function of the distance from the disk midplane, $z$. 
The geometry of the outflow and the employed coordinate system is illustrated in Supplementary Figure \ref{fig-outflow}4. 
All the uncertainties for outflow parameters are estimated by those of the fitting of PV diagrams. 
We note that the fitting procedures discussed above are too simplified and hence, further modeling including multi-transition radiative transfer calculations are required to discuss derived parameters and rotation curves more quantitatively. 

\subsection*{Data availability}

The data that support the plots within this paper and other findings of this study are available from the corresponding author upon reasonable request. 


\vspace*{12pt}

\begin{description}
\item[Supplementary Information] is linked to the online version of the paper.
\item[Acknowledgement] We are grateful to R. L. Plambeck, Y. Oya, N. Sakai, and S. Yamamoto for valuable discussions. 
This letter makes use of the following ALMA data: \\
ADS/JAO.ALMA\#2013.1.00048.S. 
ALMA is a partnership of ESO (representing its member states), NSF (USA) and NINS (Japan), together with NRC (Canada), NSC and ASIAA (Taiwan), and KASI (Republic of Korea), in cooperation with the Republic of Chile. The Joint ALMA Observatory is operated by ESO, AUI/NRAO and NAOJ. 
We thank the staff at ALMA for making observations and reducing the data. 
TH is supported by the MEXT/JSPS KAKENHI Grant Numbers 21224002, 24684011, 25108005, and 15H03646, and the ALMA Japan Research Grant of NAOJ Chile Observatory, NAOJ-ALMA-0006, 0028, 0066. 
KM is supported by the MEXT/JSPS KAKENHI Grant Number 15K17613. 
MH is supported by the MEXT/JSPS KAKENHI Grant Numbers 24540242 and 25120007. 
Data analysis were in part carried out on common use data analysis computer system at the Astronomy Data Center, ADC, of NAOJ. 
\item[Author Contributions] T.H. led the project as a principal investigator of ALMA observations and performed data analysis. 
M. M. and Y. M. interpreted the ALMA results from the theoretical point of view. 
K. M. analyzed part of the ALMA data and checked the results. 
N. M., R. A. B., and M. H. contributed to writing the manuscript. 
All the authors discussed the results and commented on the manuscript. 
\item[Author Information] The authors declare that they have no competing financial interests. \\
Correspondence and requests for materials should be addressed to T. Hirota (email: \\
tomoya.hirota@nao.ac.jp).

\end{description}

\clearpage

\section*{Supplementary Figures}

\begin{figure}[hbt]
\begin{center}
\includegraphics[width=17cm]{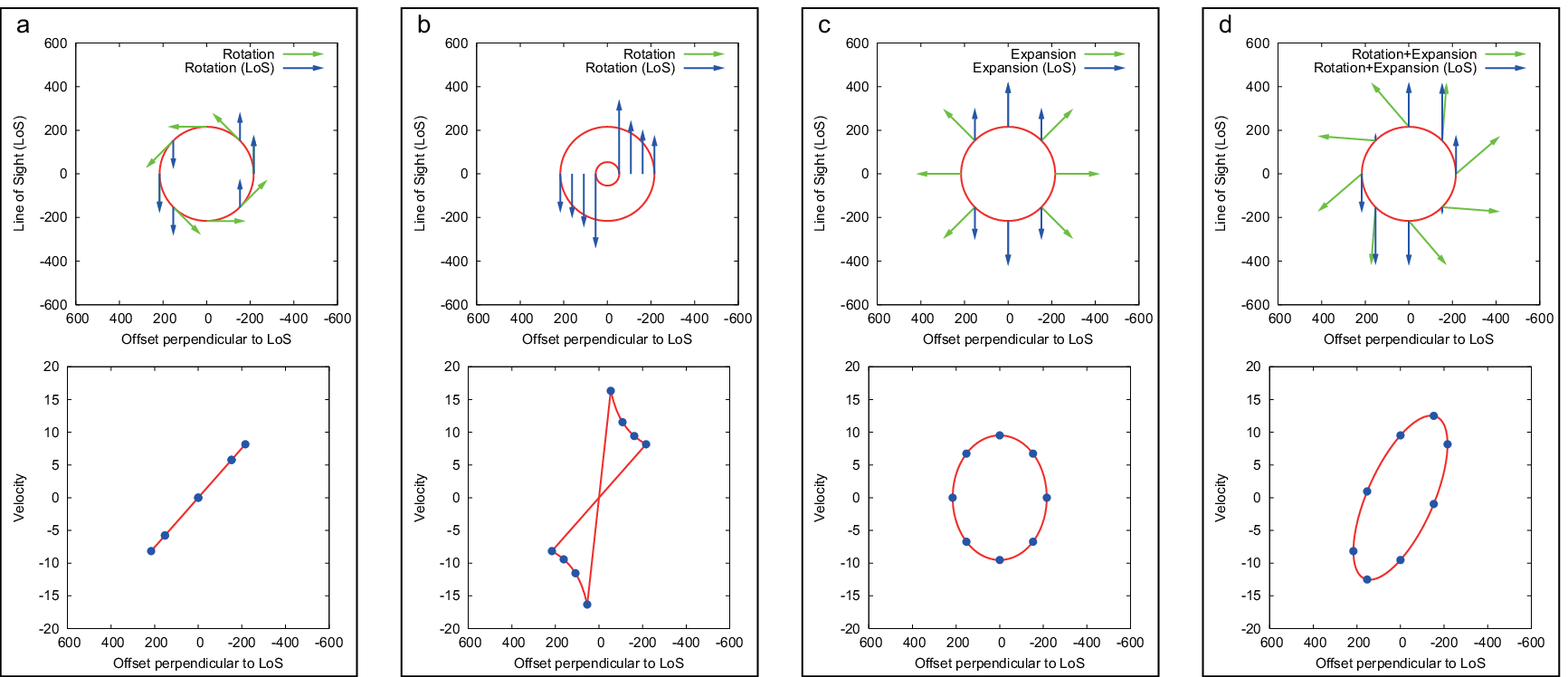}
\label{fig-model}
\end{center}
{\bf{Supplementary Figure 1:}} Model PV diagrams for the slices of the edge-on rotating and radially expanding outflow or disk. 
Top panels show the face-on view of the slices of the outflow/disk and bottom panels show corresponding PV diagram. 
An observer is supposed to be located toward the lower side of the disk in the face-on view. 
(a) Rotating ring, (b) rotating disk with a small central hole assuming the rotation curve defined by a power-law function, (c) radially expanding ring, and (d) rotating expanding ring models. 
Green and blue arrows indicate the velocity vectors and its line-of-sight components, respectively. 
\end{figure}

\clearpage

\begin{figure}[th]
\begin{center}
\includegraphics[width=17cm]{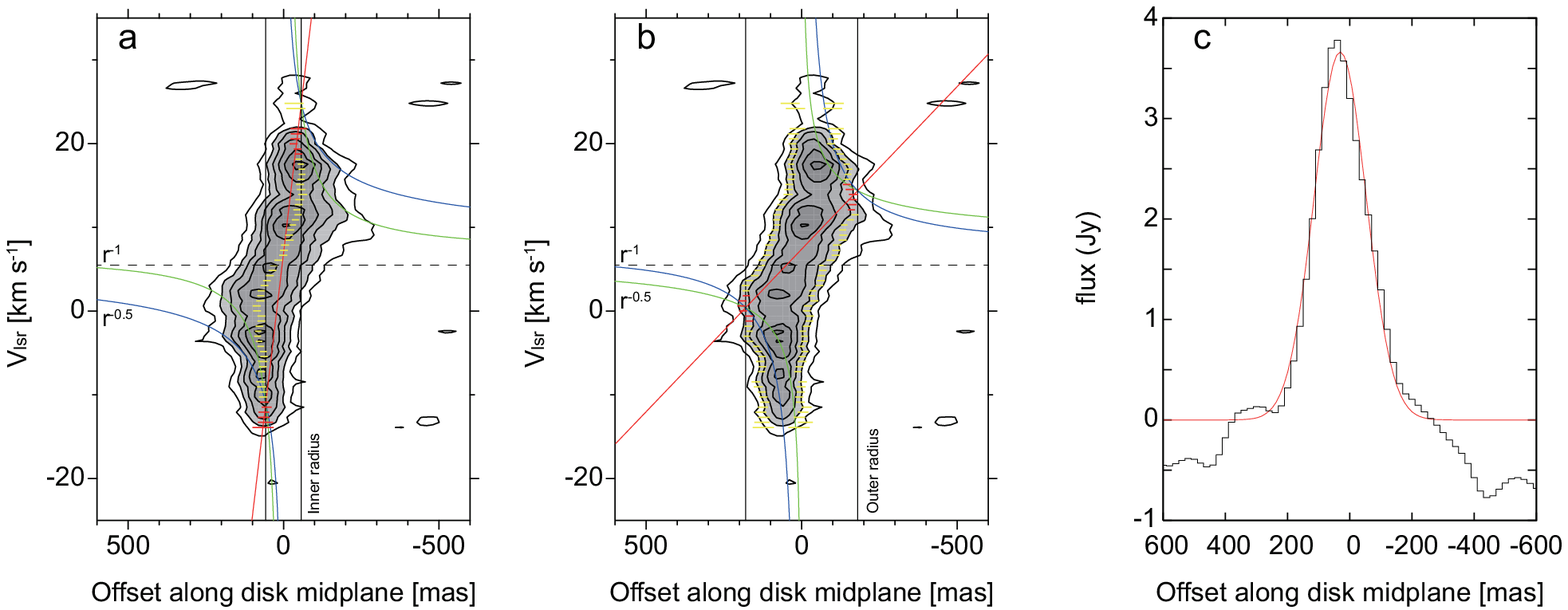}
\label{fig-sample}
\end{center}
{\bf{Supplementary Figure 2:}} Fitting to the PV diagram of of the 484~GHz Si$^{18}$O line in the disk midplane, $z$=0~au. 
(a) Linear function fitting to estimate the inner radius of the outflow lobe, $R_{in}$ (see Supplementary Figures 1(a)(b)). 
A red solid line shows the best fit model for the liner velocity gradient across the inner radius, as indicated by the solid black lines. 
Data points obtained from the Gaussian fitting of the intensity profile at each velocity channel (panel c) are plotted with error bars (1$\sigma$). 
Red symbols represent those at the highest and lowest velocity channels with the width of 3~km~s$^{-1}$ (6 channels) used for the fitting, while yellow ones represent those not used in the fitting. 
Blue and green lines represent the rotation curves corresponding to $v_{\phi} \propto r^{-0.5}$ (Keplerian rotation) and $v_{\phi} \propto r^{-1}$, respectively, assuming the same disk inner radius and rotation velocity obtained by the best-fit linear velocity gradient (i.e. not the best fit results). 
The central mass of the Keplerian rotation curve corresponds to 8.7$M_{\odot}$. 
(b) Same as the panel (a) but for the fitting of the outer radius, $R_{out}$. 
The fitting is done by using the data points around the largest radius with the velocity widths of 3~km~s$^{-1}$ (6 channels) indicated by the red symbols.  
(c) A cut of the PV diagram at $v_{lsr}$=5.5~km~s$^{-1}$, as indicated by dashed lines in panels (a) and (b). 
A red line shows the best-fit single Gaussian intensity profile. 
\end{figure}

\clearpage

\begin{figure}[th]
\begin{center}
\includegraphics[width=17cm]{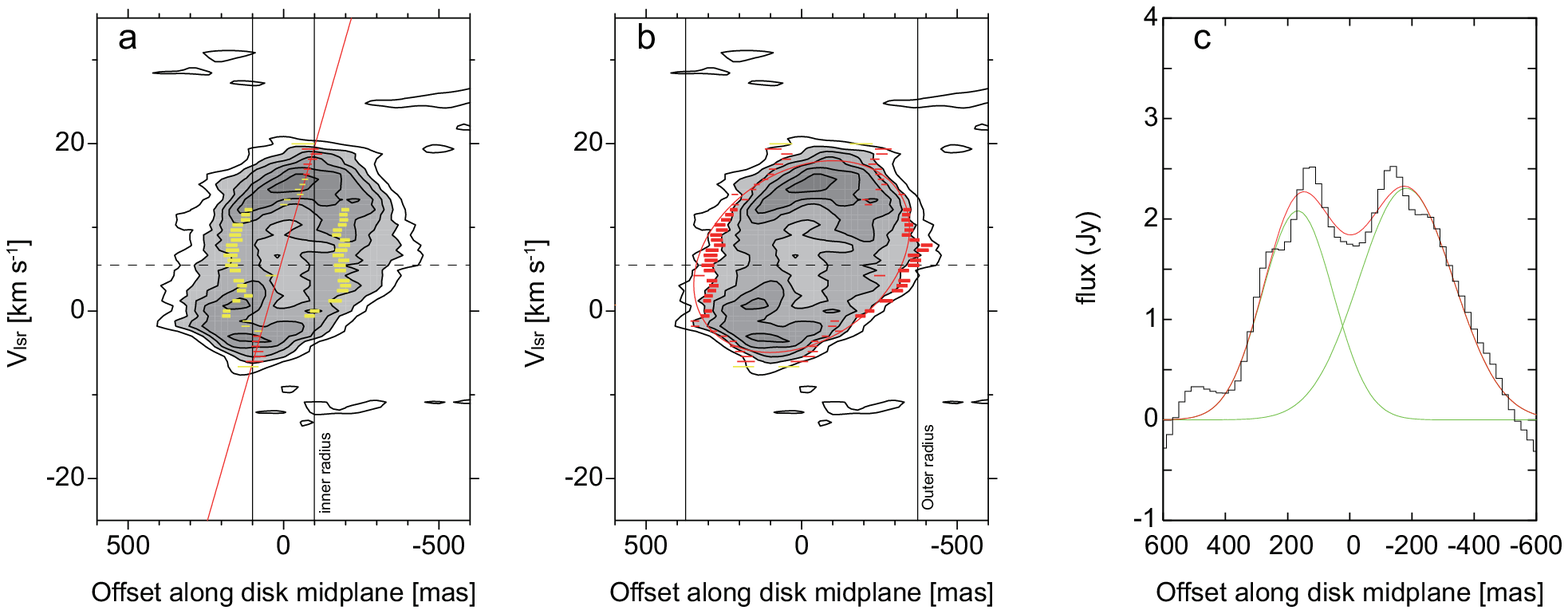}
\label{fig-sample2}
\end{center}
{\bf{Supplementary Figure 3:}} Same as Supplementary Figure 2, but at the offset position of $z$=-100~au. 
(a) Same as Supplementary Figure 2(a). 
Note that there are some velocity channels which have double peaks (panel (c)). 
Symbols with bold and thin error bars (1$\sigma$) represent the peak positions of the intensity profiles determined by double and single Gaussian fitting, respectively. 
(b) Elliptical function fitting to estimate the outer radius (see Supplementary Figures 1(c)(d)). 
A red ellipse shows the best fit model derived by using the data points at the outer boundary of the outflow lobes. 
Symbols with bold and thin error bars (1$\sigma$) represent the FWHM of the intensity profiles determined by double and single Gaussian fitting, respectively. 
(c) Same as Supplementary Figure 2(c). 
A red line represents the best-fit double Gaussian profile consisted of two peaks shown by green lines. 
\end{figure}

\clearpage

\begin{figure}[th]
\begin{center}
\includegraphics[width=17cm]{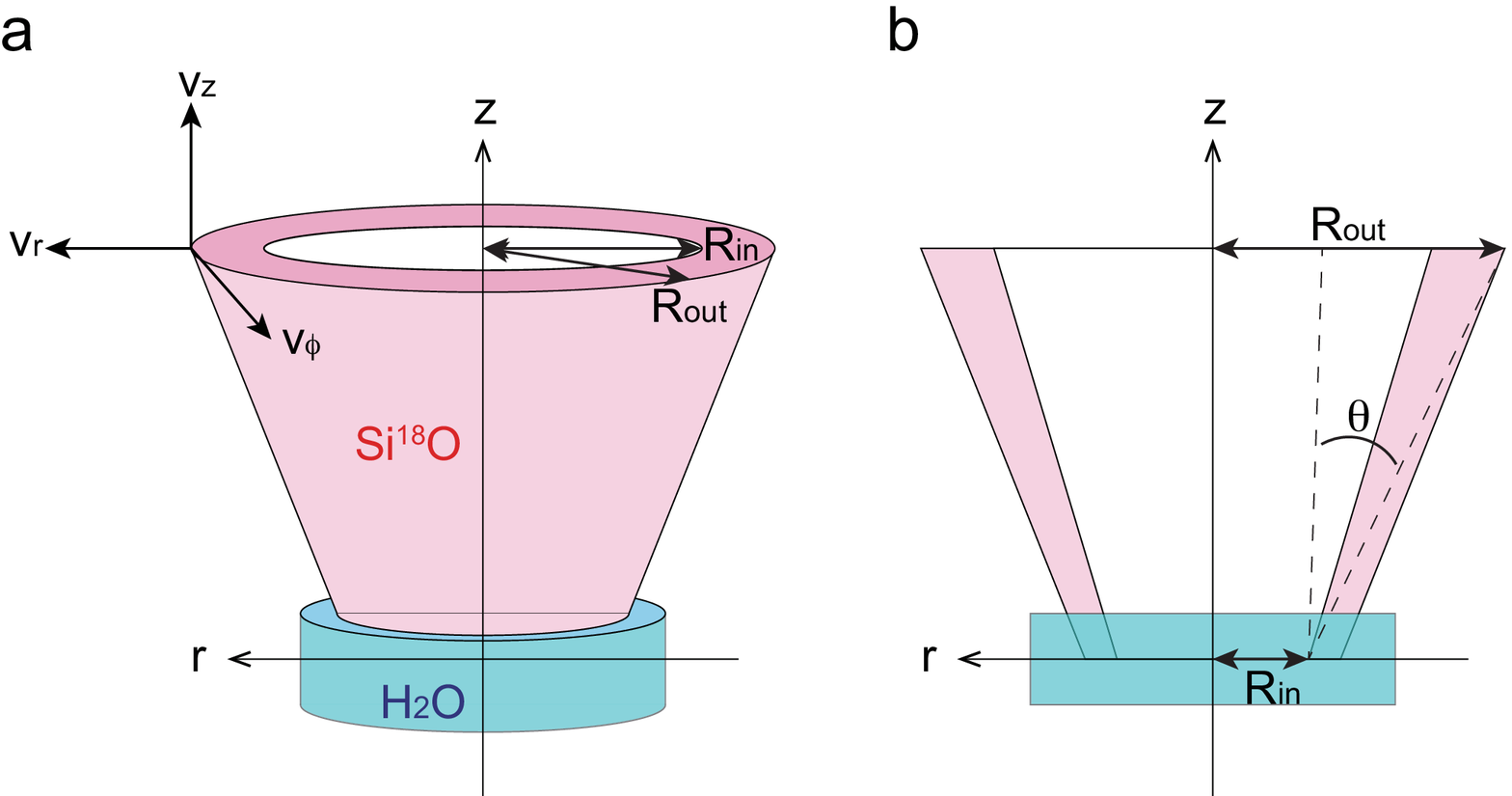}
\label{fig-outflow}
\end{center}
{\bf{Supplementary Figure 4:}} Geometry of the outflow and employed coordinate system. 
The right panel (b) shows the cross-section along the outflow axis. 
The opening angle is defined as $\theta$=$\arctan ((R_{out}(z)-R_{in}(z=0))/z)$. 
\end{figure}

\clearpage

\end{document}